\documentclass[ aip,
 reprint,]{revtex4}
\usepackage{amsmath}
\usepackage{graphicx}
\usepackage{dcolumn}
\usepackage{bm}
\usepackage[utf8]{inputenc}
\usepackage[T1]{fontenc}
\usepackage{mathptmx}
\usepackage{etoolbox}

\makeatletter
\def\@email#1#2{ \endgroup
 \patchcmd{\titleblock@produce}
  {\frontmatter@RRAPformat}
  {\frontmatter@RRAPformat{\produce@RRAP{*#1\href{mailto:#2}{#2}}}\frontmatter@RRAPformat}
  {}{}
}

\begin{document}

\title[Article Title]{Dynamics of the vortex line density in superfluids under thermal activation. \\
}
\author{Sergey Nemirovskii}
\affiliation{Institute of Thermophysics, Lavrentyev ave., 1, 630090, Novosibirsk, Russia\\
Novosibirsk State University, Pirogova 2, 630090, Novosibirsk, Russia}
\email{nemir@itp.nsc.ru}
\date{\today }

\begin{abstract}
The study explores the development of the
vortex line density in superfluids under thermal activation. This problem 
is of interest to both applied and fundamental
research, and has been investigated by many authors in various aspects.
Despite the important and impressive results obtained, a significant part of
the process, namely the kinetics of processes leading to equilibrium state,
remained unexplored. In this article, we conduct a study of kinetic
phenomena and focus our attention on the evolution of the vortex line
density (VLD) $\mathcal{L(}t\mathcal{)}$, the total length of the filament
per unit volume. The initial development of VLD is due to random thermal
fluctuations. The increase in the vortex line length $\mathcal{L(}t\mathcal{)%
}$ can be obtained based on the famous Novikov-Furutsu theorem, which shows
that the growth rate of $\mathcal{L(}t\mathcal{)}$ is proportional to a
random force correlator. As the length of the vortex filaments increases,
the interaction between the vortices becomes significant and affects the
dynamics process. At this point, we turn to the phenomenological
Feynman-Vinen theory, which offers various models for the evolution of the
quantity $\mathcal{L(}t\mathcal{)}$. Next we examine the evolution of a
vortex tangle as a combination of growth due to random thermal excitations
and decay in the Feynman-Vinen theory. Several applications leading to
significant and remarkable results are considered.
\end{abstract}

\maketitle





\nopagebreak

\section{Introduction and scientific background\protect\bigskip}

\label{Intro}

The theory of concentrated vortices occupies an important place in the study
of the hydrodynamics of both classical and quantum fluids \cite%
{Alekseenko2007}. One particular area of interest is the investigation of
stochastic ensembles of concentrated vortices. These include point vortices
in two dimensions and vortex filaments and vortex sheets in three
dimensions. The study of these ensembles of stochastic vortices has both
practical and fundamental significance and has been studied by many authors
in various aspects.

Perhaps this activity began with Onsager's work on the dynamics of point
vortices in a liquid film. Thanks to his elegant mathematical findings, this
work has received significant development, leading to the creation of a
robust mathematical framework. (See, e.g., \cite{Onsager1949a},\cite%
{lundgren1977statistical},\cite{Novikov1975},\cite{Grigor'ev1986},\cite%
{Geshev1986}, etc.). At the same time, similar studies were conducted on
superfluid helium (see , e.g., \cite{Donnelly1991},\cite{Tilley1990} and
references therein). Of course, the main advances were made in the
two-dimensional space, where the Berezinskii--Kosterlitz--Thouless
transition was predicted and studied in detail. \cite{Kosterlitz2017}.

As for the 3D case, the situation here is far from being a more or less
developed theory. Nevertheless, there are numerous works that use various
approximations. There are also scientific groups that study chaotic vortices
in both classical and quantum fluids (\cite{Alekseenko2007}, \cite%
{Bessaih2005},\cite{TimothyD.Andersen2014},\cite{Flandoli2001}, and others).
3D Quantum vortices are also intensively studied (see e.g. \ \cite%
{Copeland1991}, \cite{Williams1999},\cite{Lund1990},\cite{Antunes1998},\cite%
{Chorin1994}\cite{Jou2011a}). A special place is taken by the so-called
theory of quantum turbulence, which states that the statistical behavior of
three-dimensional vortices is very close in nature to classical turbulence
in ordinary liquids. There is no possibility to describe all these numerous
studies. It is appropriate to refer readers to the review \cite%
{Nemirovskii2013}.

Here we would like to add that the study of the stochastic dynamics of
quantum vortices is in many ways similar to the study of stochastic
topological defects, such as linear defects in solids or cosmic strings \cite%
{Kleinert1991},\cite{Kleinert1990a},\cite{Vilenkin1994}, \cite%
{Burakovsky2000}.

Despite many significant and impressive results achieved, there is still an
open question regarding the kinetics of the processes leading to an
equilibrium state. In principle, this problem could be potentially solved by
using the Fokker-Planck equation, which describes the dynamics of vortex
filaments (see, \cite{Nemirovskii2016Vns},\cite{Nemirovskii2022},\cite%
{Nemirovskii2024}). However, this approach faces great mathematical
difficulties and it is unlikely to be overcome in the near future. In this
paper, we choose a different strategy and conduct a study on kinetic
phenomena based on the Langevin approach. We focus our attention on the
evolution of the vortex line density (VLD) $\mathcal{L(}t\mathcal{)}$, which
is the total length of the vortex lines per unit volume. The initial
development of VLD occurs due to random thermal fluctuations. The increase
in the length of the vortex lines $\mathcal{L(}t\mathcal{)}$ can be obtained
based on the famous Novikov-Furutsu theorem. This shows that the growth rate
of VLD $\partial \mathcal{L(}t\mathcal{)}/\partial t$ is proportional to the
random force correlator. As the length of the vortex filaments increases,
their interaction becomes significant and affects the dynamics process. It
is impossible to fully describe this situation using the Fokker-Planck
kinetic equation because of its complexity. At this point, we turn to the
phenomenological theory of Feynman-Vinen, which offers various models for
the evolution of VLD $\mathcal{L(}t\mathcal{)}$. We analyze various
approaches to the Vinen's equation and consider the development of a vortex
tangle as a combination of random thermal excitations and VLD decay in
Feynman-Vinen theory.

In fact, the number of characteristics of a vortex tangle, which is a set of
discrete vortex filaments, is very large (formally infinite). Other
characteristics of the vortex tangle are occasionally studied in the
literature. The closest examples include the anisotropy of the tangle,
curvature of lines and other structural factors (see Schwartz's work).
However, starting with Feynman's and Vinen's works, it is generally believed
that vortex line density (VLD), $\mathcal{L(}t\mathcal{)}$, is the most
significant and important characteristic. Many aspects, such as the time for
quantum turbulence to develop, can be examined from the perspective of $%
\mathcal{L(}t\mathcal{)}$ dynamics. In this context, it is quite reasonable
to use the evolution of vortex line density to characterize the kinetic
development of a quantum vortex tangle. This will not cause any
misunderstandings.

\section{Langevin and Fokker-Planck equation}

\label{FP}

We consider the Langevin dynamics of vortex loops in a three-dimensional
space without boundaries. The equation of motion for the vortex line
elements reads

\begin{widetext}
\begin{equation}
\mathbf{\dot{s}=\dot{s}}_{B}+\mathbf{v}_{s}+\alpha \mathbf{s}^{\prime }(\xi
)\times (\mathbf{v}_{ns}-\mathbf{\dot{s}}_{B})+\alpha ^{\prime }\mathbf{s}%
^{\prime }(\xi )\times \mathbf{s}^{\prime }(\xi )\times (\mathbf{v}_{ns}-%
\mathbf{\dot{s}}_{B})+\mathbf{f(}\xi _{0},t).  \label{Master eq}
\end{equation}
\end{widetext}

Here $\mathbf{s(}\xi \mathbf{,}t\mathbf{)}$ is the radius-vector of the
vortex line points; $\xi $ is a label parameter, which in this case
coincides with the arc length; $\mathbf{\dot{s}}_{B}$ is the self induced,
Biot-Savart, propagation velocity of the vortex filament at a point $\mathbf{%
s}$, determined by the Biot-Savart law; $\mathbf{v}_{ns}$ is the relative
velocity between the normal and superfluid components of helium II; $\mathbf{%
s^{\prime }}$ is the derivative with respect to the arc length, $\alpha $
and $\alpha ^{\prime }$ are the friction coefficients that describe the
interaction of the vortex filament with the normal component, and $\mathbf{f(%
}\xi ,t\mathbf{)}$ is the Langevin force. The Langevin force is assumed to
be a white noise with the following correlator
\begin{equation}
F=\left\langle \mathbf{f}_{i}(\xi _{1},t_{1})\mathbf{f}_{j}(\xi
_{2},t_{2})\right\rangle \ =D_{1}\delta _{ij}\delta \ (t_{1}-t_{2})\delta \
(\xi _{1}-\xi _{2}).  \label{FDT}
\end{equation}%
Here $i$ and $j$~are the spatial components; $t_{1}$and $t_{2}$ are the
arbitrary time moments; $\xi _{1\text{ }}$and $\xi _{2}$ denote any points
on the vortex line, $\rho _{s}$ is the density of the superfluid component; $%
\kappa $ is the quantum of circulation equal to $2\pi (\hbar /m)$. The
renormalized \ diffusion coefficient $D_{1}$ is related to temperature and
to the problem parameters as $D_{1}=\frac{2k_{B}T\alpha }{\rho _{s}\kappa }$
with dimension m$^{3}$/s. It differs from the usual diffusion coefficient $D$
(with dimension m$^{2}$/s) which is used in numerical simulation of the
stochastic vortex filaments by a factor of $1/a$. We will use these two
quantities interchangeably in our calculations.

Quantity $a$ is the length parameter of the model. For a cubic lattice
model, it is the cube edge, for polymer chains it is the elementary step. In
the case of quantum vortices, the quantity $a$ is the healing (or coherence)
length and can be taken as the core size (see for details, e.g., \cite%
{Kleinert1991},\cite{Kleinert1990a}, \cite{Copeland1991}). In the discrete
version, which is used for numerical simulations, the initial step along the
line $\Delta \xi $ can be chosen as $a.$

The term with $\alpha ^{\prime }$\ does not enter into the fluctuation
dissipation theorem. Its role is reduced to the fact, that it simply changes
the self-induced velocity (Biot-Savart term). It could be said that for the
dissipation of energy only the $\alpha $\ term is responsible, while the
reactive mutual-friction $\alpha ^{\prime }$\ just slightly renormalizes the
inertial term in conventional hydrodynamics. This situation (for different
reasons) was discussed earlier in the paper \cite{Finne2003}.

Given the value of $\mathbf{s}(\xi ,t)$ at the initial time, the Langevin
equation generates a time-dependent probability distribution functional $%
\mathcal{P}(\{\mathbf{s}(\xi )\},t)$ \ \cite{Nemirovskii2004} for the
stochastic vector $\mathbf{s}(\xi ,t)$, which can be formally written as:%
\begin{equation}
\mathcal{P}(\{\mathbf{s}(\xi )\},t)=\left\langle \delta \left( \mathbf{s}%
(\xi )-\mathbf{s}(\xi ,t)\right) \right\rangle .  \label{pdf}
\end{equation}%
Here $\delta $ is the delta functional in the space of configurations of
vortex loops. \ An averaging was performed over an ensemble of random forces.

The Fokker-Planck equation for the time evolution of the quantity $\mathcal{P%
}(\{\mathbf{s}(\xi )\},t)$ can be derived from the equation of motion (\ref%
{Master eq}) in a standard way (see e.g.\cite{Zinn-Justin2002},\cite%
{Nemirovskii2004})

\begin{widetext}
\begin{equation}
\frac{\partial \mathcal{P}}{\partial t}+\int d\xi \frac{\delta }{\delta
\mathbf{s}(\xi _{1})}{\Huge \{}\left[ \mathbf{\dot{s}}_{B}+\mathbf{v}%
_{s}+\alpha \mathbf{s}^{\prime }(\xi )\times (\mathbf{v}_{n}-\mathbf{v}_{s}-%
\mathbf{\dot{s}}_{B})\right] \mathcal{P}+\frac{1}{2}\left\langle \mathbf{f}%
_{i}(\xi _{1},t_{1})\mathbf{f}_{j}(\xi _{2},t_{2})\right\rangle \frac{\delta
}{\delta \mathbf{s}(\xi _{2})}{\Huge \}}\mathcal{P}\ =0.  \label{FP1}
\end{equation}
\end{widetext}

In Ref. \cite{Nemirovskii2016Vns} it was shown that a set of vortex
filaments in superfluids under the action of a random Langevin force in the
presence of a counterflow with a relative velocity $\mathbf{v}_{ns}$ come
into a state of thermodynamic equilibrium with the Gibbs distribution. The
properties of this distribution, as well as important physical consequences,
were studied in a series of works by the author \cite{Nemirovskii2016Vns},%
\cite{Nemirovskii2022},\cite{Nemirovskii2024}.

\section{Evolution of the vortex line length due to thermal activation.}

\label{dynamics copy(4)}

As noted earlier, the Fokker-Planck equation method, in principle, can serve
as a basis for studying of non-stationary problems. However, this path is
technically unreachable due to the lack of a solution to the Fokker-Planck
equation in the general case and is therefore hardly acceptable. Therefore,
here we will choose another strategy and will consider the problem based on
the Langevin approach initially. In this article we will limit ourselves to
studying the evolution of the vortex line density $\mathcal{L(}t\mathcal{)}$.

We start with the evolution of an arbitrary element with a length $\delta l$%
. The evolution of this linear element $\delta \mathbf{l}$\textbf{\ }(it is
a vector!)\textbf{\ }in a fluid flow with velocity $\mathbf{v(r,}t)$
satisfies the following relation (see, e.g., \cite{Batchelor1953},\cite%
{Schwarz1978}):

\begin{equation}
\frac{d\delta \mathbf{l}}{dt}=(\delta \mathbf{l\cdot \nabla )v+}o(\left\vert
\delta \mathbf{l}\right\vert ).  \label{dlover dt}
\end{equation}

Let's demonstrate how the equation for the evolution of the vortex line
density $\mathcal{L}(t)$\ can be derived (see paper by Schwarz \cite%
{Schwarz1978}). To calculate $\partial \mathcal{L}/\partial t,$ we use the
relation for the rate of a change in the length $\partial \delta l/\partial
t $ \ of element $\delta \mathbf{l}$. Assuming that the label variable $\xi $
is not exactly equal to the arc length, we have $\delta l=\left\vert \mathbf{%
s}^{\prime }\right\vert $ $\delta \xi .\ $Then\ the following chain of
equations takes place:%
\begin{equation}
\frac{\partial \delta l}{\partial t}=\frac{\partial \left\vert \mathbf{s}%
^{\prime }\right\vert \delta \xi }{\partial t}=\frac{\left\vert \mathbf{s}%
^{\prime }\right\vert }{\left\vert \mathbf{s}^{\prime }\right\vert }\frac{%
\partial \left\vert \mathbf{s}^{\prime }\right\vert \delta \xi }{\partial t}=%
\frac{\mathbf{s}^{\prime }}{\left\vert \mathbf{s}^{\prime }\right\vert }%
\frac{\partial \mathbf{s}^{\prime }\delta \xi }{\partial t}=\mathbf{s}%
^{\prime }\dot{\cdot \mathbf{s}}^{\prime }\delta \xi .  \label{dl/dt}
\end{equation}%
From Eq. (\ref{dl/dt}) it follows, in particular, that if there is an
external force $\mathbf{f(}\xi ,t)$ acting on a line (see Eq. \ref{Master eq}%
), then the growth (or reduction) of the length is equal to $\mathbf{s}%
^{\prime }\mathbf{f}^{\prime }$. In the last equality we have returned to
the condition that the label variable $\xi $ coincides with the arc length,
i.e. $\left\vert \mathbf{s}^{\prime }\right\vert =1.$ If we (temporarily)
let go of the regular terms in Eq. (\ref{Master eq}) other than thermal
fluctuations, then Eq. (\ref{dl/dt}) can be rewritten as:

\begin{equation}
\frac{\partial \delta l}{\partial t}=\mathbf{s}^{\prime }\cdot \mathbf{f}%
^{\prime }(\xi ,t)\delta \xi .  \label{dl due f}
\end{equation}

This expression can be treated using the well known properties of Gaussian
variables - the Novikov-Furutsu theorem (see, e.g. , \cite{Zinn-Justin2002}).

\begin{equation}
\ \left\langle \mathbf{s}^{\prime }\cdot \mathbf{f}^{\prime }(\xi
,t)\right\rangle =\int d\xi _{1}\frac{\delta \mathbf{s}_{i}^{\prime }(%
\mathbf{\xi },t)}{\delta \mathbf{f}_{j}^{\prime }(\xi _{1},t)}\left\langle
\mathbf{f}_{j}^{\prime }(\xi _{1},t)\mathbf{f}_{i}^{\prime }(\xi
,t)\right\rangle \   \label{Novikov  relation}
\end{equation}

The correlation function included in the \ref{Novikov relation} can be
obtained by spatial differentiation of the delta function appearing in the
fluctuation dissipation theorem \ref{FDT}. Assuming that there is a lower
limit $\ a$ for spatial correlations, we can write $\left\langle \mathbf{f}%
^{\prime }(\xi _{1},t)\mathbf{f}^{\prime }(\xi ,t)\right\rangle $ as%
\begin{equation}
\left\langle \mathbf{f}_{i}^{\prime }(\xi _{1},t)\mathbf{f}_{j}^{\prime
}(\xi _{2},t)\right\rangle =\frac{1}{a^{2}}D_{1}\delta _{ij}\delta \
(t_{1}-t_{2})\delta \ (\xi _{1}-\xi _{2}).  \label{FDT f'f'}
\end{equation}

The Green function $\delta \mathbf{s}^{\prime }(\mathbf{\xi },t)/\delta
\mathbf{f}^{\prime }(\xi _{1},t)$ when $t\rightarrow t^{\prime }$ can be
found from the following considerations. From $\dot{\mathbf{s}}^{\prime }=%
\mathbf{f}^{\prime }(\xi ,t)$ it follows that
\begin{equation*}
\mathbf{s}^{\prime }(\mathbf{\xi },t)=\int_{0}^{t}\theta (t-t^{\prime \prime
})\mathbf{f}^{\prime }(\xi ,t^{\prime })dt^{\prime \prime }.
\end{equation*}%
From it, we have the functional derivative
\begin{equation*}
\frac{\delta \mathbf{s}_{i}^{\prime }(\mathbf{\xi },t)}{\delta \mathbf{f}%
_{j}^{\prime }(\xi _{1},t^{\prime })}=\delta _{ij}\theta (t-t^{\prime \prime
}).
\end{equation*}%
According to Zihn-Justin \cite{Zinn-Justin2002}, we can do the following
steps. Smearing the $\theta $ in the vicinity of zero, and supposing that
the smeared function to be $\theta (x)-1/2$ is odd, we get $\theta
(t-t)=1/2. $\ Gathering everything, and using the fluctuation dissipation
theorem \ref{FDT f'f'} we obtain (after dividing by the volume of the
system) the following relation for the growth of the VLD under the influence
of thermal activation.

\begin{equation}
\frac{\partial \mathcal{L}}{\partial t}=\int d\xi _{1}\frac{1}{2}\frac{%
2k_{B}T\alpha _{f}}{\rho _{s}\kappa a}\frac{1}{a^{2}}=\mathcal{L}\frac{D_{1}%
}{2a^{3}}=\mathcal{L}\frac{D}{2a^{2}}.  \label{L via D}
\end{equation}

Here the integration is performed on a unit volume. The equation (\ref{L via
D}) is the key final formula of Section III, devoted to the growth of vortex
filaments under the influence of thermal disturbances. It follows naturally
from all the relationships given in this Section.

It follows from the Eq. (\ref{L via D}) that under the influence of thermal
fluctuations the value of VLD $\mathcal{L}(t)$ increases monotonically.
Obviously, there must be a mechanism stop this process. Unfortunately, the
Fokker-Planck method is not suitable for solving this problem, as it is not
stationary and equilibrium, and it cannot be solved in analytic form at the
moment. We plan to use a phenomenological theory called the Feynman-Vinen
theory, to describe the macroscopic dynamics of superfluid turbulence. This
theory is based on certain assumptions and experimental data. We will apply
it to describe the dynamics of vortices under thermal excitations. In the
Feynman-Vinen theory there are some contradictory points of view and various
versions of the final equations. Therefore, it is appropriate to outline the
main points before moving on to discussing \ the main issue of combining
this theory with a method based on thermal fluctuations. This macroscopic
theory on VLD evolution is presented in Appendix A.

\section{Applications. Development of a quantum vortex tangle in various
conditions.}

\subsection{Velocity-free case. Steady case}

\label{dynamics}

Let's combine the results of Section III on the development of a quantum
vortex tangle due to thermal activation with the Feynman-Vinen model of
evolution VLD $\mathcal{L}(t)$ as described in Appendix A. First \ we will
study the velocity-free case. This case with zero velocity is very
interesting\textit{, }because it relates to an important issue of nucleation
of the vortex loops due to thermal activation (\cite{Iordanskii1965},\cite%
{Langer1967}, \cite{Donnelly1991}). Another advantage is that a stationary
version of this problem was considered in the author's article based on the
equilibrium approach (see, paper \cite{Nemirovskii2024}). It is interesting
to compare two different approaches

Combining Eq. (\ref{L via D}) related to the thermal growth of the tangle
and the Vinen equation following from the Feynman-Vinen theory, stated in
Appendix A, we write the evolution equation for VLD $\mathcal{L}(t)$ in the
following form:%
\begin{equation}
\frac{d\mathcal{L}(t)}{dt}=-\beta _{un}\mathcal{L}^{2}+\mathcal{L}\frac{D_{1}%
}{2a^{2}}.  \label{VE un}
\end{equation}

We have introduced the universal coefficient $\beta _{un}$ to emphasize that
there are multiple models for the macroscopic theory of VLD evolution (see
Appendix A). Remembering that the coefficient $D_{1}=\frac{2k_{B}T\alpha }{%
\rho _{s}{s}\kappa }\ $with dimension$\ $(m$^{3}$/s), we get the stationary
solution VLD $\mathcal{L}_{st}$ is

\begin{equation}
\mathcal{L}_{st}=\frac{1}{\beta _{un}}\frac{1}{2a^{3}}D_{1}  \label{VLD st}
\end{equation}

The result (\ref{VLD st}) is obtained based on a combined thermal activation
considerations with the Feynman-Vinen theory. Of course, it is very tempting
to compare this with the result based on the equilibrium state theory. The
corresponding relation follows from the partition function calculated
earlier by the author \cite{Nemirovskii2016Vns},\cite{Nemirovskii2022},\cite%
{Nemirovskii2024}. The comparison is presented in Fig 1. The upper line was
obtained from the formula (\ref{VLD st}), with the coefficient $\beta _{un}$%
, taken from Schwartz's works \cite{Schwarz1978},\cite{Schwarz1988}. The
lower line was obtained using the Gibbs distribution see details in paper
\cite{Nemirovskii2024}) for more information.

\begin{figure}[tbp]
\includegraphics[width=7cm]{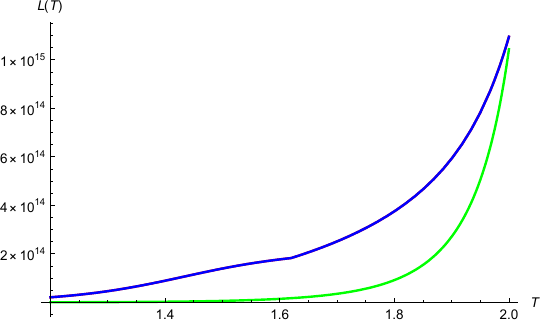}
\caption{(Color online) Fig. 1. The vortex line density \ $\mathcal{L}(T)$ \
(1/cm$^{2}$) as a function of temperature $T$. The lower line is obtained
from the partition function for a thermodynamically equilibrium ensemble of
vortex filaments, as described in the paper \protect\cite{Nemirovskii2024}.
The upper line was calculated based on the combined kinetic consideration
plus the Feynman-Vinen theory, as described in Sections \protect\ref%
{dynamics copy(4)}-\protect\ref{dynamics copy(3)}, (see Eq. (\protect\ref%
{VLD st}).}
\label{Fig1}
\end{figure}

The result obtained above seems to be very significant and remarkable.
Indeed, the value of \ VLD $\mathcal{L}(T)$ is obtained in two completely
different ways. One way is based on the Gibbs distribution. Note that the
Gibbs distribution is a universal tool; there are many works on the study of
stochastic dynamics that start with the Gibbs distribution. The other way is
based on a combination of the Langevin-type kinetic equation and the
Vinen-Feynman phenomenological theory (See Appendix A of the presented
paper).

The accuracy of the results depends on the method used in each case. The
macroscopic theory of the VLD is an extremely inaccurate theory, that
depends on many parameters. These parameters give quite different values in
various experiments, numerical results and theories. In contrast, the method
based on the Gibbs distribution is more accurate, because it depends on
fewer parameters that are defined much more precisely. Apparently, this
circumstance leads to different curves in Figure 1. Nevertheless, the result
seems is very remarkable. Indeed, two completely different approaches lead
to values that are quite close in order of magnitude. This is an indirect
indication of the validity of both approaches.

\subsection{Velocity-free case. Unsteady case}

\label{dynamics copy(1)}

Let us turn to non-stationary phenomena. The development of the density of
vortex lines obeys Eq. (\ref{VE un}). When solving this equation, a problem
with the divergence at the lower limit immediately arises, that is, when VLD
$\mathcal{L}(t=0)$ tends to zero. Therefore, the question of the initial
conditions arises. Here we are faced with two circumstances. First, the
solution of the equation is very insensitive to the initial level, as
calculations show. Even changes by several orders of magnitude have
virtually no effect on the result. Second, a similar problem arises in the
classical Vinen equation (\ref{VE}). The question arises is either about the
initial conditions or about background values of VLD $\mathcal{L}_{back}$
and value of VLD $\mathcal{L}_{back}$ of the order of $100-1000$ 1/cm$^{2}$
which are used in the calculations.

\bigskip
\begin{figure}[tbp]
\includegraphics[width=7cm]{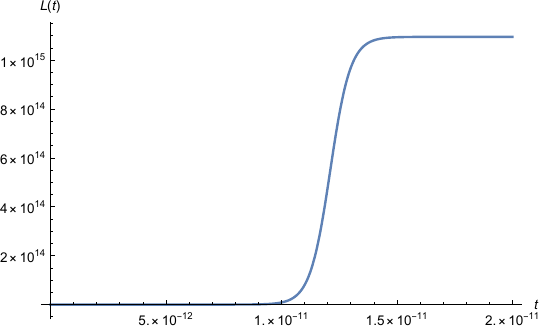}
\caption{(Color online) Fig. 2. The vortex line density \ $\mathcal{L}(t)$
as a function of time $t$ (s). The asymptotic value of VLD $\mathcal{L}%
(t\rightarrow \infty )$ is close to the values obtained earlier in steady
state (see Fig 1). The time to reach the equilibrium value is very short,
which implies that the vortex tangle adapts to the temperature change almost
instantly.}
\end{figure}

Figure 2 shows the time dependence of the VLD $\mathcal{L}(t)$ at the
initial condition of $\mathcal{L}_{back}$ $=1000$ 1/cm$^{2}$. The asymptotic
value of the VLD $\mathcal{L}(t\rightarrow \infty )$ corresponds to the
equilibrium value obtained previously in two different ways (see Fig 1).
This indirectly confirms the validity of our reasoning. The second feature
is that the time it takes to reach the equilibrium value is very short. This
means that the vortex tangle adapts to the temperature change almost
instantly.

\subsection{Velocity-free case. Numerical simulations}

\label{dynamics copy(2)}

The formulation of the above problem using a discrete approach is very
important for numerical considerations. Numerical simulation is the main
tool for solving the stochastic dynamics of quantum vortex filaments subject
to random force, due to its complexity.

A discrete version of the formalism developed above can be obtained from the
following considerations (see for details \cite{Nemirovskii2020e}) We
represent a vortex filament as a set of discrete points (\textquotedblleft
beads\textquotedblright ) $\mathbf{s}_{m}$ ($m$ varies from $1$ to $N$)
initially separated by a distance $a_{d}$.

The correspondence between continuous and discrete descriptions can be
represented by a set of rules, exposed in \cite{Nemirovskii2020e}.

Let us introduce the following discrete version of the probability
distribution functional $\mathcal{P}(\{\mathbf{s}_{m},t)=\left\langle
\mathcal{P}^{M}\right\rangle =\left\langle \delta \left( \mathbf{s}_{m}-%
\mathbf{s}_{m}(t)\right) \right\rangle .$The evolution of $\mathcal{P}(\{%
\mathbf{s}_{m},t)$ can be presented in the following form: Compare with the
continuum version of the Fokker-Planck equation \ref{FP1}

\begin{eqnarray}
&&\frac{\partial \mathcal{P}}{\partial t}+\sum_{m}\frac{\delta }{\delta
\mathbf{s}_{m,\eta _{1}}}\left\{ \alpha \overset{\cdot }{\mathbf{s}}_{m,\eta
_{1}}\mathcal{P}\right\} +  \label{FP discrete} \\
&&+\frac{1}{2}\sum_{m,n}\left\langle \mathbf{\zeta }_{\eta _{1}}\mathbf{(}%
m,t_{1})\mathbf{\zeta }_{\eta _{2}}\mathbf{(}n,t_{2})\right\rangle \frac{%
\delta }{\delta \mathbf{s}_{n,\eta _{1}}}\frac{\delta }{\delta \mathbf{s}%
_{m,\eta _{2}}}\mathcal{P}\ =0,\ \ \   \notag
\end{eqnarray}

The discrete version of the fluctuation-dissipation theorem has the
following \ form

\begin{equation}
\left\langle \mathbf{\zeta }_{\eta _{1}}\mathbf{(}m,t_{1})\mathbf{\zeta }%
_{\eta _{2}}\mathbf{(}n,t_{2})\right\rangle ~=D~\delta _{mn}~\delta
(t_{1}-t_{2})\delta _{\eta _{1}\eta _{2}}.  \label{FDT discrete 1}
\end{equation}

Note that, unlike the continuous version, the intensity of random forces is
determined by a coefficient $D$ with the dimension m$^{2}$/s. The relation (%
\ref{FDT discrete 1}) allows us to formulate the problem in terms of the
diffusion coefficient, without involving the issue of temperature. This
circumstance is extremely important for numerical studies. The
\textquotedblleft temperature\textquotedblright\ in such situations is an
artificial quantity that has nothing to do with real temperature. In this
sense, we study a model problem that, however, has many important
applications including superfluid turbulence.

Now we will consider from a theoretical point of view one of the works \cite%
{Kondaurova2003}, which is dedicated to the numerical modelling of chaotic
vortex filaments. The simulation was carried out at the temperature of $1.07$
K, (which corresponds to the friction coefficient of $\alpha =0.0098$), the
steps along the vortex line was $a_{n}=2\pi \cdot 10^{-5}$ cm, and the
diffusion coefficient was $D_{n}=4\ast 10^{-4}$ cm$^{2}$/s. The results of
the calculations are shown in Fig. 3 where the value of the VLD $\mathcal{L}%
(t)$ is depicted.

\begin{figure}[tbp]
\includegraphics[width=7cm]{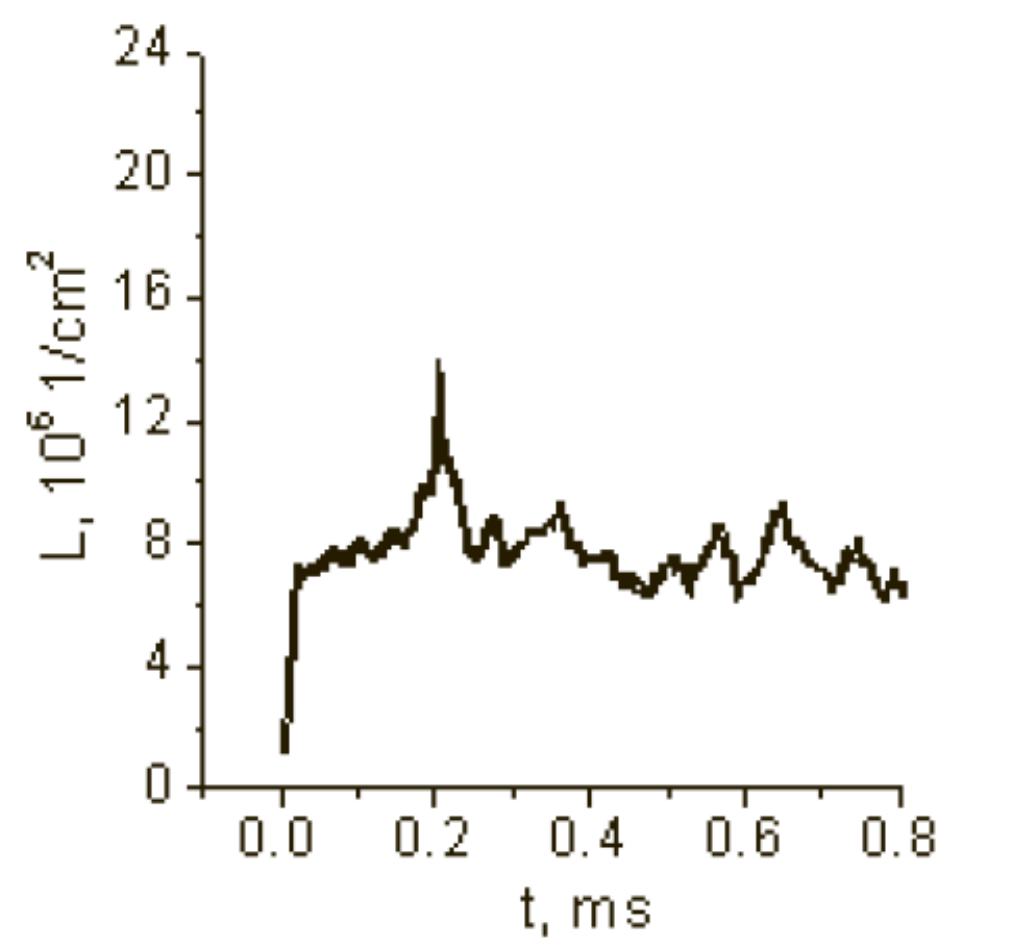}
\caption{(Color online) Fig. 3. Plots of the t vortex lines density $%
\mathcal{L}(t)$ as a function of time.$t$ This curve is the result of
numerical simulation in paper \protect\cite{Kondaurova2003}. Initially, the
vortex line density increases to a steady state. After a certain value of
the density has \ been archived, many small loops develop and the VT begins
to decay as a result of their reconnections.}
\end{figure}

This curve is the result of numerical simulation in paper \cite%
{Kondaurova2003}. Initially, the density of vortex lines increases until it
reaches a steady state. After a certain value of the density has \ been
archived, many small loops develop and the vortex tangle starts to decay.

Let's examine what the kinetic theory developed in this paper suggests. The
Eq. (\ref{VE un}) with parameters from the numerical work \cite%
{Kondaurova2003} leads to a solution for the time dependence of VLD $%
\mathcal{L}(t)$. This is shown in Fig. 4.

\begin{figure}[tbp]
\includegraphics[width=7cm]{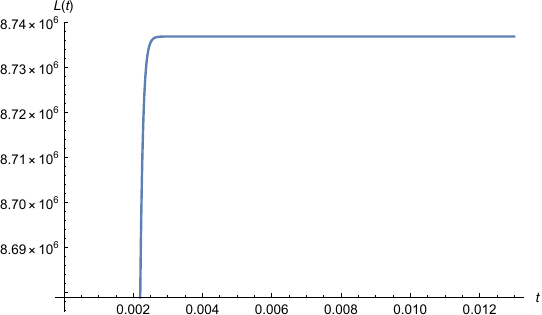}
\caption{(Color online)Fig. 4. Plots of the t vortex lines density $\mathcal{%
L}(t)$ (1/cm$^{2}$)as a function of time $t$ (s), obtained in this paper
based on Eq. (\protect\ref{VE un}) with the remarks, nade in Sec. \protect
\ref{dynamics copy(2)}. It is seen that the asymptotic value of VLD $%
\mathcal{L}(t\rightarrow \infty )$ is close to the value, observed in
experimental work \protect\cite{Kondaurova2003}. }
\end{figure}

It is seen that the temporal behavior qualitatively coincides with that
presented in paper \cite{Kondaurova2003}. In particular, the asymptotic
value of the VLD $\mathcal{L}(t\rightarrow \infty )$ matches the value,
observed in experimental work \cite{Kondaurova2003}.\ The second feature is
that the time required \ to reach the equilibrium value is also close to
that observed in the work \cite{Kondaurova2003}. This indicates that the
theory developed above functions effectively.

\subsection{The counterflow case}

\label{counterflow}

The case with a non-zero counterflow velocity is very interesting. As it was
discussed in subsection \ref{dynamics copy(3)} , the classical form of the
generating term in the Vinen-Schwarz equation (\ref{VE}) or (\ref{VE un})
can be modified using the so called alternative equation (\ref{alternate VE}%
). In steady state these two equations yield identical results, but in
unsteady situations they produce very similar results, which does not allow
us to select a more appropriate form.

Let us consider this situation from the position of the theory of thermal
excitation of vortex lines, which was developed above. From our formalism it
follows that these two equations, when supplemented by random thermal
influences, lead to very different results even in the steady state. Below
we work with a renormalized diffusion coefficient $D_{1}=\frac{2k_{B}T\alpha
}{\rho _{s}\kappa }$ with dimension $\frac{m^{3}}{s}$. In order to
distinguish between these two cases, we introduce the variables $\mathcal{L}%
_{Vin}$ and $\mathcal{L}_{alt}.$The corresponding equations are:

\begin{equation}
\frac{\partial \mathcal{L}_{Vin}}{\partial t}=\alpha _{V}\left\vert
v_{ns}\right\vert \mathcal{L}_{Vin}^{3/2}-\beta _{V}L\mathcal{L}_{Vin}^{2}+%
\mathcal{L}_{Vin}\frac{D}{2a^{3}},  \label{CounterVE}
\end{equation}%
\begin{equation}
\frac{\partial \mathcal{L}_{alt}}{\partial t}=a_{alt}v_{ns}^{2}\mathcal{L}%
_{alt}-\beta _{alt}\mathcal{L}_{alt}^{2}+\mathcal{L}_{alt}\frac{D}{2a^{3}}.
\label{counter alt VE}
\end{equation}%
For small counterflow velocities $v_{ns}$, the solutions take the following
forms%
\begin{equation}
\mathcal{L}_{Vin}=\allowbreak \frac{1}{2a^{3}\beta _{V}}D+\frac{1}{2a^{6}}%
v_{ns}\frac{\alpha _{V}}{\beta _{V}^{2}}\sqrt{2a^{9}\beta _{V}D}+\frac{1}{2}%
v_{ns}^{2}\frac{\alpha _{V}^{2}}{\beta _{V}^{2}}  \label{L Vin}
\end{equation}%
\begin{equation}
\mathcal{L}_{alt}=\allowbreak \frac{1}{2a^{3}}\frac{D}{\beta _{alt}}%
+v_{ns}^{2}\frac{a_{alt}}{\beta _{alt}}  \label{L alt}
\end{equation}

The structure of the solutions for $\mathcal{L}_{Vin}$ and $\mathcal{L}%
_{alt} $ looks like this: there is a terms that does not depend on velocity $%
v_{ns} $ and coincide with the stationary solution for VLD $\mathcal{L}(t)$
(Equation (\ref{VLD st})). From the results of paragraph IV.A it follows
that this value of $\mathcal{L}(t)$ corresponds to the Gibbs equilibrium
distribution, which does not depend on velocity. The remaining terms in
these solutions are related to velocity $v_{ns}$ and represent the
quantities that researchers of superfluid turbulence work with. The last
terms in both parts of equations (\ref{L Vin}) and (\ref{L alt}) completely
coincide with the classical solutions of Feynman Vinen theory, which can be
found in Appendix A. However, the classical form (Eq. (\ref{L Vin}))
introduces an additional term compared to those, discussed earlier. This
term has a linear relationship between the vortex line density $\mathcal{L}%
_{Vin}$ and the applied velocity. To our knowledge, among the huge number of
works devoted to the study the dependence $\mathcal{L(}v_{ns}\mathcal{)}$,
there has been only one work, where such a dependence was observed. This
work has been heavily criticized (see \cite{Buttke1987} and citation within).

Thus, we \ can argue that from our Langevin formalism, it follows that the
classical form of the Vinen equation leads to an incorrect (linear)
dependence of the vortex line density on the applied velocity. Consequently,
we can conclude that the alternative form of the equation for the evolution
of the vortex line density, $\mathcal{L}(t)$ is more accurate.

\section{Discussion and conclusion}

In this article, we conducted a study on the kinetic behavior of stochastic
ensembles of quantum vortices using the Langevin method. We focused our
attention on the evolution of the vortex line density, $\mathcal{L(}t%
\mathcal{)}$ - the total length of the filaments per unit volume. Initially,
the VLD develops due to random thermal fluctuations. The increase in the
vortex line length $\mathcal{L(}t\mathcal{)}$ can be determined using the
famous Novikov-Furutsu theorem, which shows that the growth rate of the VLD $%
\partial \mathcal{L(}t\mathcal{)}/\partial t$ is proportional to the random
force correlator.

As the vortex filaments become longer, their interactions become more
noticeable and affect the dynamics of the system. To better \ understand
this process better, we turn to the phenomenological Feynman-Vinen theory.
This theory offers different models for how the vortex tangle forms. We then
explore how the vortex tangle evolves as a result of both growth due to
random thermal energy and decay, according to the Feynman-Vinen theory. The
applications of this research have produced important and remarkable
outcomes, demonstrating the practical value of this field of study.

So, the result on the VLD $\mathcal{L}(T)$ shown in Fig. 1 is quite
remarkable. Indeed, the value of \ VLD $\mathcal{L}(T)$ is obtained using
two completely methods. One method is based on the Gibbs distribution, an
universal tool in the stochastic dynamics. The other method is based on a
Langevin-type equation developed here.\ The similarity between the results
obtained confirms the validity of our chosen approach. The same conclusion
applies to the unsteady situation, described in Section \ref{dynamics
copy(1)}\

The results of the numerical simulations, presented in Section \ref{dynamics
copy(2)}, are also noteworthy. Numerical methods play a crucial role in
studying the stochastic dynamics of quantum vortex filaments subjected to
random forces. The close agreement between the experimental and theoretical
findings validates our approach.

Finally, the case of a nonzero counterflow velocity , explored in Sec. \ref%
{counterflow} is undoubtedly very significant and remarkable. Indeed, the
results of this section demonstrate that from our Langevin formalism it
follows that the classical form of the Vinen's equation (\ref{VE}) predicts
an incorrect linear dependence of the vortex line density on the applied
velocity. At the same time, an alternative form of the equation (\ref%
{alternate VE}) for the time evolution of the vortex line density $\mathcal{L%
}(t)$ provides more reasonable results.

\bigskip

The study was financially supported by the Russian Science Foundation (Grant
No. 23-22-00128).

\renewcommand{\theequation}{\U{413}\U{40f}.\arabic{equation}} %
\renewcommand{\theequation}{\thesection.\arabic{equation}} %
\setcounter{equation}{0}\appendix

\section{Macroscopic theory of the VLD evolution. Feynman-Vinen scenario}

\label{dynamics copy(3)}

The term "superfluid turbulence" (ST) or quantum turbulence (QT) was
introduced by Feynman in his seminal paper \cite{Feynman1955}, where he
explained the results of Gorter and Mellink (see \cite{Gorter1949}), who
observed a sharp increase in the temperature difference in a counterflowing
He-II when the velocity exceeded a certain, rather small value. At first
glance, this seemed to be a question of the end of the phenomenon of
superfluidity (or the associated super thermoconductivity). However, Feynman
associated the Gorter-Mellink crisis with the appearance of a disordered set
of quantized vortex lines, or vortex tangles (VT), which exert resistance on
the flow of the normal component that carries entropy.

Feynman suggested that in a helium flow, for example, in a counterflow, when
a certain critical relative velocity $\mathbf{v}_{ns}=\mathbf{v}_{n}-\mathbf{%
v}_{s}$ is exceeded, quantized vortices appear, the same as in the case of
rotating helium. The causes and processes of vortex formation were not
discussed. Unlike the case of rotating helium, where due to symmetry the
generated vortex filaments are aligned along the rotation axis, the vortex
lines in the counterflow are arbitrarily oriented and tangled. By analogy
with ordinary liquids, Feynman suggested that: "The vortex lines twist in an
increasingly complex manner, increasing their length due to the kinetic
energy of the main flow. That is, the vortex lines form a tangled vortex
tangle with a large length of vortex filaments. A complex irregular velocity
field is added to the uniform velocity of motion, the energy for increasing
the length of the filaments and for the velocity fluctuations is supplied
from the main flow."

As the length of the lines increases, they fill the volume of the liquid
more densely, and the processes of interaction between the vortex lines
begin to play an increasingly important role. Feynman suggested that as the
density of vortex filaments increases, they often collide in space, and the
result of this collision is the reconnection of the lines. The reconnection
of vortex filaments leads to the processes of merging and decay
(recombination) of loops. Feynman suggested that the latter property
dominates, that is, on average, the decay of vortex loops occurs. This leads
to an avalanche-like process of forming smaller and smaller loops. When the
size of small rings reaches the order of interatomic distances, which is the
final stage of the cascade, the vortex motion degenerates into thermal
excitations (rotons?).

Feynman's qualitative model was further developed in the classic works of
Vinen \cite{Vinen1957}, \cite{Vinen1957a,Vinen1957b}. He formulated these
ideas in quantitative terms and, in particular, derived an equation (which
bears his name) that describes the macroscopic dynamics of a vortex tangle.
Vinen's equation describes the evolution of the total length of vortex lines
per unit volume, $\mathcal{L}(t)$.\
\begin{equation}
\frac{d\mathcal{L}}{dt}=a_{V}|\mathbf{v}_{ns}|\mathcal{L}^{3/2}-\beta _{V}%
\mathcal{L}^{2}.  \label{VE}
\end{equation}%
In parallel, Vinen conducted an experiment to verify the validity of the
formula (\ref{VE}) and determine the values {}{}of the phenomenological
parameters $a_{V}$ and $\beta _{V}$. However, due to the scatter in the
experimental data and the experimental error, Vinen noted that another form
of the $d\mathcal{L}/dt$ dependence as a function of $\mathcal{L}$ and $%
\mathbf{v}_{ns}$ cannot be ruled out. In particular, he acknowledged that
the experimental result (although perhaps worse) may correspond to a
different dependence of the generating term leading to the so-called
"alternative" equation (see also \cite{Nemirovskii1995})
\begin{equation}
\frac{d\mathcal{L}}{dt}=a_{alt}\mathbf{v}_{ns}^{2}\mathcal{L}-\beta _{alt}%
\mathcal{L}^{2}.  \label{alternate VE}
\end{equation}

Schwarz \cite{Schwarz1978},\cite{Schwarz1988} made the next significant
contribution to the derivation of the VLD evolution equation from
microscopic considerations. He conducted the direct numerical simulation of
superfluid turbulence in a counterflowing He II induced by the heat load,
and based on the obtained results, developed a semi-quantitative theory on
the structure of the vortex tangle. This allowed the analytical derivation
of the Vinen equation

\begin{equation}
\frac{d\mathcal{L}}{dt}=a_{s}|\mathbf{v}_{ns}|\mathcal{L}^{3/2}-\beta _{s}%
\mathcal{L}^{2}.  \label{SE}
\end{equation}%
The Schwarz Eq. (\ref{SE}) has exactly the same form as the Vinen equation (%
\ref{VE}). However, firstly, the coefficients $\alpha _{s},\beta _{s}$
differ quite strongly (by about half an order of magnitude) from each other.
Secondly, and perhaps more importantly, the physical meaning of the equation
(\ref{SE}) is quite different from the proposed Feynman-Vinen scenario. In
particular, the second decay term, which is responsible for reducing the VLD
$\mathcal{L}(t)$, is also related to the dynamics of the line (See. Eq. (\ref%
{Master eq})) and depends on the friction coefficient. That is, although
reconnection processes were used in the numerical modeling of the vortex
tangle structure, the equation for the changing the line length itself did
not take into account the Feynman and Vinen mechanisms associated with loop
fragmentation due to reconnections.

This mechanism was taken into account in the works of the author \cite%
{Nemirovskii2006},\cite{Nemirovskii2008}. In these works, the evolution of a
vortex tangle is considered from the perspective of the kinetics of vortex
loops with a Brownian structure. Based on the study of the balance
Boltzmann-type equation, the problem of the distribution of vortex loops $%
n(l)$ in the space of their sizes was solved.

This solution is not thermodynamically equilibrium, but on the contrary, it
describes a state with two mutual fluxes of length (or energy) in the of
vortex loop sizes. The term \textquotedblleft flux\textquotedblright\ simply
\ means the redistribution of length (or energy) among the loops of
different sizes due to reconnections.

Depending on the temperature, either the mechanism of vortex loop
enlargement or the mechanism of their fragmentation (described by Feynman)
predominates. In either case, this affects the evolution and must be
included in a equation of the type of Vinen (\ref{VE}) or Schwarz (\ref{SE}%
). The final expression for the pure "flow" of length in the space of loop
sizes

\begin{equation}
P_{net}=P_{+}-P_{-}=C_{F}\kappa \mathcal{L}^{2}  \label{Fluxes eq}
\end{equation}%
We named the constant $C_{F}$ in honor of Feynman who was the first to
discuss the evolution of the vortex line density due to the reconnection
processes.

Collecting all the contributions to $d\mathcal{L}(t)/dt$ from both the
deterministic and collisional processes, and taking into account the
expression for the net \textquotedblright flux\textquotedblright\ (\ref%
{Fluxes eq}), we finally have:%
\begin{equation}
\frac{d\mathcal{L}(t)}{dt}=a_{V}|\mathbf{v}_{ns}|\mathcal{L}^{3/2}-\beta _{V}%
\mathcal{L}^{2}-\left\vert C_{F}\right\vert \kappa \allowbreak \mathcal{L}%
^{2}.  \label{ViEN}
\end{equation}%
The equation for the evolution of $\mathcal{L}(t)$ (\ref{ViEN}) includes
both a regular part and a stochastic part related to the length flux due to
reconnections. \ref{Fluxes eq}.

Many of the details described in the appendix can be found in the author's
reviews \cite{Nemirovskii1995}, \cite{Nemirovskii2013}.


\end{document}